\documentclass[twocolumn,prl,superscriptaddress,showpacs,letter]{revtex4}

\usepackage{graphicx}
\usepackage{amssymb}
\usepackage{amsfonts}

\DeclareGraphicsExtensions{.eps}

\begin{document}

\title{Polaritonic Bistability in Semiconductor Microcavities}

\date{\today}

\author{A. Baas}
\affiliation{Laboratoire Kastler Brossel, Universit\'{e} Paris 6, Ecole
Normale Sup\'{e}rieure et CNRS,\\
UPMC Case 74, 4 place Jussieu, 75252 Paris Cedex 05, France}

\author{J. Ph. Karr}
\affiliation{Laboratoire Kastler Brossel, Universit\'{e} Paris 6, Ecole
Normale Sup\'{e}rieure et CNRS,\\
UPMC Case 74, 4 place Jussieu, 75252 Paris Cedex 05, France}

\author{E. Giacobino}
\affiliation{Laboratoire Kastler Brossel, Universit\'{e} Paris 6, Ecole
Normale Sup\'{e}rieure et CNRS,\\
UPMC Case 74, 4 place Jussieu, 75252 Paris Cedex 05, France}

\begin{abstract}
We report the observation of polaritonic bistability in semiconductor microcavities in the strong
coupling regime. The origin of bistability is the polariton-polariton interaction, which gives rise to a
Kerr-like nonlinearity. The experimental results are in good agreement with a simple model taking
transverse effects into account.
\end{abstract}


\maketitle

\section{Introduction}

In high finesse semiconductor microcavities with embedded quantum wells, the demonstration of the strong
coupling regime between the quantum well excitons and the cavity photons \cite{weisbuch} has opened the
way to a refined manipulation of a new species, cavity polaritons, that are mixed light-matter
eigenstates. While strong coupling or normal mode coupling appears for very small photon numbers,
polaritons exhibit a number of nonlinear behaviors \cite{khitrova}. Polariton bleaching has been observed
at high excitation density \cite{houdre} and is predicted to give rise to optical bistability
\cite{tredicucci}. With intermediate excitation densities for which strong coupling still exists, these
systems exhibit strong nonlinear emission, due to parametric polariton amplification. The nonlinearity
comes from the exciton part of the polariton through coherent exciton-exciton scattering.

The polariton scattering must fulfil phase matching conditions for the in-plane wave vector of the
considered polaritons. If $\mathbf{k_{P}}$ is the wave vector of the excitation, two polaritons with wave
vector $\mathbf{k_{P}}$ scatter to give polaritons with wave vectors $\mathbf{0}$ and $\mathbf{2k_{P}}$.
In addition, energy conservation implies that $E(0)+E(2k_{p})=2 \; E(k_{P})$. Most of the experiments on
the polariton parametric amplification have been performed in the magic angle configuration
\cite{magic,nonres}, where $k_{P}$ is the nontrivial solution for the energy conservation condition.
However, there also exists a trivial solution $\mathbf{k_{P}}=0$ where only the $\mathbf{k}=0$ mode is
involved. In this case, in the same way as in degenerate parametric amplification, polaritonic wave
mixing gives rise to phase dependent amplification \cite{messin} and eventually to polariton squeezing
\cite{karr03}.

In this geometry of excitation, the effective Hamiltonian (at first order) for the polariton-polariton
interaction \cite{ciuti1,ciuti2} is analogous to the Hamiltonian of an optical Kerr medium. The
difference is that the refraction index depends on the polariton number instead of the photon number.
This gives rise to a bistable behavior for high enough excitation intensities, in the same way as for a
Kerr medium in a cavity. Optical bistability has already been observed in quantum well microcavities in
the weak coupling regime \cite{oudar} ; it has been predicted to occur in the strong coupling regime due
to exciton bleaching \cite{tredicucci}, however in different conditions.

In this paper we give experimental evidence for a bistable behavior in a microcavity sample in the strong
coupling regime. To our knowledge, this is the first observation of bistability in the strong coupling
regime. We also investigate the nonlinear patterns that appear in the transverse plane. We show that the
main features of the experimental results can be explained satisfactorily by treating the
polariton-polariton interaction at first order, i.e. in terms of a polariton Kerr effect.

The paper is organized as follows. In section I, we give the effective Hamiltonian for the polariton
system, using the same set of hypotheses that has been used for the study of the "magic angle"
configuration \cite{ciuti1,ciuti2}. We establish the evolution equation for the polariton field, and we
solve the steady-state regime in order to compute the bistability threshold, as well as the reflectivity
and transmission spectra. Section II is the experimental study of the bistability regime. We show that it
is necessary to take transverse effects into account and we compare the experimental results with the
prediction of the model studied in section I, including a simple treatment of the transverse effects.

\section{I. Model}

\subsection{Hamiltonian}

The linear Hamiltonian for excitons and cavity photons is $H=\sum\limits_{\mathbf{k}} H_{\mathbf{k}}$
with

\begin{eqnarray}
H_{\mathbf{k}}&=& E_{exc}\left( k\right) b_{\mathbf{k}}^{\dagger } b_{\mathbf{k}}+E_{cav}\left( k\right)
a_{\mathbf{k}}^{\dagger }a_{\mathbf{k}} \\ &+& \frac{\Omega_{R}}{2}\left( a_{\mathbf{k}}^{\dagger
}b_{\mathbf{k}}+b_{\mathbf{k}}^{\dagger }a_{\mathbf{k}}\right) \nonumber
\end{eqnarray}

In this equation $a_{\mathbf{k}}$ and $b_{\mathbf{k}}$ are the creation operators for photons and
excitons with a wave vector $\mathbf{k}$ in the layer plane. Because of the translational invariance in
the cavity plane, photons can only interact with excitons having the same $\mathbf{k}$. $E_{cav}\left(
k\right)$ ($E_{exc}\left( k\right)$) is the cavity (exciton) dispersion and $\Omega_{R}$ is the Rabi
interaction energy between excitons and photons. The normal modes of the linear Hamiltonian
$H_{\mathbf{k}}$ are called cavity polaritons. The annihilation operators $p_{\mathbf{k}}$,
$q_{\mathbf{k}}$ for the lower and upper polaritons are given by

\begin{eqnarray}
p_{\mathbf{k}} &=& X_{k} b_{\mathbf{k}} - C_{k} a_{\mathbf{k}} \label{def1} \\
q_{\mathbf{k}} &=& C_{k} b_{\mathbf{k}} + X_{k} a_{\mathbf{k}} \label{def2}
\end{eqnarray}

where $X_{k}$ and $C_{k}$ are the Hopfield coefficients

\begin{eqnarray}
X_{k} &=& \left( \frac{\delta _{k}+\sqrt{\delta _{k}^{2}+\Omega_{R} ^{2}}}{2\sqrt{%
\delta _{k}^{2}+\Omega_{R} ^{2}}} \right)^{1/2} \label{hopfield1}\\
C_{k} &=& \left( \frac{\Omega_{R} ^{2}}{2\sqrt{\delta _{k}^{2}+\Omega_{R} ^{2}}\left( \delta
_{k}+\sqrt{\delta _{k}^{2}+\Omega_{R} ^{2}}\right)} \right)^{1/2} \label{hopfield2}
\end{eqnarray}

with $\delta _{k}=E_{cav}(k) -E_{exc}(k)$. In the polariton basis, the linear Hamiltonian reads

\begin{equation}
H_{\mathbf{k}} = E_{LP}(k) p_{\mathbf{k}}^{\dagger} p_{\mathbf{k}} + E_{UP}(k) q_{\mathbf{k}}^{\dagger}
q_{\mathbf{k}}
\end{equation}

where $E_{LP}(k)$ ($E_{UP}(k)$) is the lower (upper) polariton dispersion, given by

\begin{equation}
E_{LP(UP)}(k) = E_{exc}(k) + \frac{\delta_{k}}{2} -(+) \frac{1}{2} \sqrt{\delta_{k}^{2} + \Omega_{R}^{2}}
\end{equation}

The Coulomb interaction between the carriers gives rise to two additional terms: an effective
exciton-exciton interaction term and an anharmonic saturation term in the light-exciton coupling. The
exciton-exciton interaction term is

\begin{equation}
H_{exc-exc}=\frac{1}{2}\sum\limits_{\mathbf{k},\mathbf{k'},\mathbf{q}}V_{q}
b_{\mathbf{k}+\mathbf{q}}^{\dagger} b_{\mathbf{k'}-\mathbf{q}}^{\dagger} b_{\mathbf{k}}b_{\mathbf{k'}}
\end{equation}

with $V_{q}\simeq V_{0}=6e^{2}a_{exc}/\epsilon _{0}A$ provided $qa_{exc}\ll 1$, $a_{exc}$ being the
two-dimensional Bohr radius of the exciton, $\epsilon _{0}$ the dielectric constant of the quantum wells
and $A$ the quantization area. The saturation term writes

\begin{eqnarray}
&& H_{sat}= \\ &&-\sum\limits_{\mathbf{k},\mathbf{k'},\mathbf{q}} V_{sat} \left(
a_{\mathbf{k}+\mathbf{q}}^{\dagger }b_{\mathbf{k'}-\mathbf{q}}^{\dagger
}b_{\mathbf{k}}b_{\mathbf{k'}}+a_{\mathbf{k}+\mathbf{q}}
b_{\mathbf{k'}-\mathbf{q}}b_{\mathbf{k}}^{\dagger }b_{\mathbf{k'}}^{\dagger }\right) \nonumber
\end{eqnarray}

with $V_{sat}=\Omega_{R}/2n_{sat}A$, where $n_{sat}=7/\left( 16\pi a_{exc}^{2}\right)$ is the exciton
saturation density. As long as the nonlinear terms are small compared to the Rabi splitting $\Omega_{R}$,
it is possible to neglect the nonlinear interaction between the upper and lower branches, which yields
non secular terms. The two polaritons are then virtually decoupled and it is more appropriate to use the
polariton basis. In addition, we consider a resonant excitation of the lower branch by a
quasi-monochromatic laser field and we will focus our attention on the evolution of the lower branch
polariton. In terms of the lower polariton operator the Hamiltonian is $H=H_{LP} + H_{PP}^{eff}$. The
free polariton term is $H_{LP}=\sum_{\mathbf{k}} E_{LP}(k) p_{\mathbf{k}}^{\dagger} p_{\mathbf{k}}$. The
effective polariton-polariton interaction term is

\begin{equation}
H_{PP}^{eff}=\frac{1}{2}\sum\limits_{\mathbf{k},\mathbf{k^{\prime}},\mathbf{q}}
V_{\mathbf{k},\mathbf{k^{\prime}},\mathbf{q}}^{PP} p_{\mathbf{k+q}}^{\dagger }p_{\mathbf{k^{\prime
}-q}}^{\dagger }p_{\mathbf{k}}p_{\mathbf{k^{\prime}}}
\end{equation}

with

\begin{eqnarray}
&& V_{\mathbf{k},\mathbf{k^{\prime}},\mathbf{q}}^{PP} = V_{0} X_{|\mathbf{k^{\prime}-q}|} X_{k}
X_{|\mathbf{k+q}|} X_{k^{\prime}}
 \label{nleff} \\&& + 2 V_{sat} X_{|\mathbf{k^{\prime}-q}|} X_{k} \left( C_{|\mathbf{k+q}|} X_{k^{\prime}} + C_{k^{\prime}}
X_{|\mathbf{k+q}|} \right) \nonumber
\end{eqnarray}

For the typical parameters $\Omega_{R}$=3 meV, $a_{exc}$=100 \AA \ and $\epsilon _{0}$=(3.5)$^{2}$
$\epsilon _{0}^{vacuum}$ we find $V_{sat}/V_{0}\simeq$ 0.012. Therefore in equation (\ref{nleff}) we can
neglect the saturation term with respect to the Coulomb interaction term (except for the case of an
extreme negative detuning, where one can have $C_{k} \gg X_{k}$):

\begin{equation}
V_{\mathbf{k},\mathbf{k^{\prime}},\mathbf{q}}^{PP} \simeq V_{0} X_{|\mathbf{k+q}|} X_{k^{\prime}}
X_{|\mathbf{k^{\prime}-q}|} X_{k}
\end{equation}

In the following we consider a resonant excitation at normal incidence (in the $\mathbf{k}=0$ direction)
and we study the reflected field (again in the $\mathbf{k}=0$ direction). In this case the interaction of
the $\mathbf{k}=0$ polaritons with other modes gives rise to the collision broadening calculated by C.
Ciuti in Ref.~\cite{ciuti0}. He predicted a threshold behavior of the broadening. The exciton density at
threshold is $n_{exc}$ = 7.10$^{9}$ cm$^{-2}$ at $\delta$ = 3 meV for a sample with a Rabi splitting
$\Omega$ = 3 meV. Below this threshold, we can neglect the collision broadening and keep only the lowest
order term. Only a Kerr-like nonlinear term remains:

\begin{eqnarray}
H_{PP}^{eff} &=& \frac{1}{2} V_{\mathbf{0}} p_{\mathbf{0}}^{\dagger} p_{\mathbf{0}}^{\dagger}
p_{\mathbf{0}} p_{\mathbf{0}} \label{nonlineaire}
\end{eqnarray}

\begin{figure}[htbp]
\centerline{\includegraphics[clip=,width=8cm]{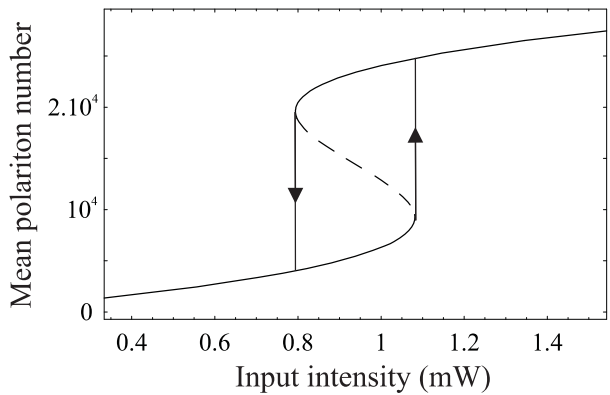}} \caption{Intensity of the polariton field
(i.e. the mean number of polaritons $n_{p}$) versus the input power $I^{in}$ in milliwatts. The nonlinear
coefficient is $V_{0}$=4.5 10$^{-5}$ meV (corresponding to a spot of 50 $\mu$m in diameter). The cavity
and exciton linewidths are $\gamma_{a}$=0.12 meV and $\gamma_{b}$=0.075 meV and the polariton linewidth
is taken as equal to $\gamma_{p}=C_{0}^{2} \gamma_{a} + X_{0}^{2} \gamma_{b}$ (see text). The
cavity-exciton detuning is $\delta$=0. The laser detuning is $\delta _{p}$=-2.5 $\gamma _{p}$. In the
bistable region, the dotted line is the unstable branch. The arrows indicate the hysteresis cycle
obtained by scanning the input power in both directions.} \label{bistability}
\end{figure}

Finally we include a term describing the coupling between the cavity mode and the external pump field of
frequency $\omega_{L}$, treated as a classical field with amplitude $A^{in}$ \cite{walls} :

\begin{equation}
H_{pump} = i \hbar \sqrt{2 \gamma_{1}} \left( A^{in} exp(-i \omega_{L} t) a^{\dagger} - h. c. \right)
\end{equation}

where $\gamma_{1}$ the dissipation coefficient of the front mirror of the cavity.

\subsection{Evolution equation}

The evolution equation for the lower polariton operator then reads :

\begin{equation}
\frac{dp}{dt} = -\left( \gamma _{p}+i\delta _{p}\right) p-i\alpha_{p} p^{\dagger }pp - C_{0}
\sqrt{2\gamma _{1}}A^{in} \label{evolpolariton}
\end{equation}

where $\delta _{p} =E_{LP}(0)/\hbar - \omega_{L}$ is the frequency detuning between the polariton
resonance and the laser. $\gamma_{p}$ is the polariton linewidth, which is given in our simple coupled
oscillator model by $\gamma_{p}=C_{0}^{2} \gamma_{a} + X_{0}^{2} \gamma_{b}$ where $\gamma_{a}$ and
$\gamma_{b}$ are the bare cavity and exciton linewidths, respectively. This gives correct values of the
linewidth at low excitation density for a limited range of detunings around zero detuning. At higher
excitation the collision broadening should be taken into account. $\alpha_{p}$ is the polariton effective
nonlinear coefficient given by

\begin{equation}
\alpha _{p}= \frac{X_{0}^{4} V_{0}}{\hbar} \label{alphapol}
\end{equation}

Equation (\ref{evolpolariton}) will be our starting point for the study of the nonlinear effects. It is
analogous to the evolution equation of the optical field in a cavity containing an ideal passive Kerr
medium. It is therefore expected that the nonlinear polariton system should exhibit bistability. Let us
note that our system is made more complex by the composite nature of the cavity polaritons. All the
parameters, like the polariton linewidth, the nonlinear coefficient and the coupling to radiation are
functions of the cavity-exciton detuning, which determines the photon and exciton contents of the
polariton.

\subsection{Steady-state regime}

We rewrite equation (\ref{evolpolariton}) for the mean fields and we solve the stationary regime:

\begin{equation}
\frac{d\left\langle p\right\rangle }{dt}= -\left( \gamma _{p}+i\delta _{p}\right) \left\langle
p\right\rangle -i\alpha _{p}n_{p}\left\langle p\right\rangle -C_{0}\sqrt{2\gamma _{1}}\left\langle
A^{in}\right\rangle = 0 \label{intracav2}
\end{equation}

where $n_{p}=\left| \left\langle p\right\rangle \right| ^{2}$ is the mean number of polaritons.
Multiplying equation (\ref{intracav2}) by its conjugate, we obtain an equation for $n_{p}$

\begin{equation}
n_{p}\left( \gamma _{p}^{2}+\left( \delta _{p}+\alpha _{p}n_{p}\right) ^{2}\right) =2\gamma
_{1}C_{0}^{2}I^{in} \label{intracav}
\end{equation}

The plot of $n_{p}$ versus the excitation power shows a bistable behavior for certain values of $\delta
_{p}$, as can be seen in Fig.~\ref{bistability}. For a range of values of the driving laser power the
polariton number is found to have two possible values, located on the higher and the lower stable
branches of the curve (the intermediate branch is well known to be unstable). The turning points are
given by the equation $\frac{dI^{in} }{dn_{p}}=0$:

\begin{equation}
3\alpha _{p}^{2}n_{p}^{2}+4\alpha _{p}n_{p}\delta _{p}+\gamma _{p}^{2}+\delta _{p}^{2}=0
\label{bistabilite}
\end{equation}

The discriminant of this equation writes

\begin{equation}
\Delta =\alpha _{p}^{2}\left( \delta _{p}^{2}-3\gamma _{p}^{2}\right)
\end{equation}

A bistable behavior is obtained for positive values of the discriminant, i.e. $\delta _{p}^{2}>3\gamma
_{p}^{2}$. Moreover the solutions for $n_{p}$ should be positive real numbers. Combining these two
conditions, bistability is obtained when

\begin{equation}
\delta _{p}<-\sqrt{3}\gamma _{p}\text{, i.e. }\omega _{L}>\omega _{p}+\sqrt{3}\gamma _{p} \label{domain}
\end{equation}

In this case, the value of $n_{p}$ corresponding to the bistability turning point writes

\begin{equation}
n_{p}^{1}=\frac{-2\delta _{p}-\sqrt{\delta _{p}^{2}-3\gamma _{p}^{2}}}{3\alpha _{p}}
\end{equation}

and one can simply obtain from (\ref{intracav}) the corresponding threshold for the excitation intensity:

\begin{eqnarray}
I^{in} &=& {\frac{-\left( 2\delta _{p}+\sqrt{\delta _{p}^{2}-3\gamma _{p}^{2}} \right)} {27\alpha
_{p}C_{0}^{2}\gamma
_{1}}} \nonumber \\
&\times& \left( \delta _{p}^{2}+3\gamma _{p}^{2}-\delta _{p}\sqrt{\delta _{p}^{2}-3\gamma
_{p}^{2}}\right)
\end{eqnarray}

\subsection{Bistability threshold}

\begin{figure}[htbp]
\centerline{\includegraphics[clip=,width=8cm]{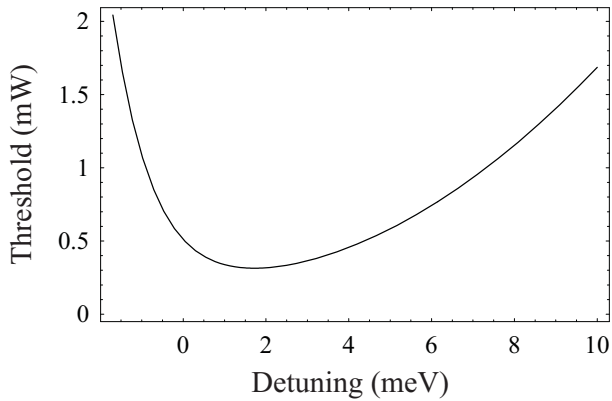}} \caption{Variations of the bistability
threshold in milliwatts versus the cavity-exciton detuning. The parameters are the same as in
Fig.~\ref{bistability}. } \label{bistab}
\end{figure}

The lowest threshold is obtained when the detuning between the laser and the polariton resonance $\delta
_{p}$ is equal to $ -\sqrt{3} \gamma_{p}$. The corresponding threshold is

\begin{equation}
I_{thr}^{in}=\frac{4\gamma _{p}^{3}}{3\sqrt{3}\alpha _{p}C_{0}^{2}\gamma _{1}} \label{threshold}
\end{equation}

It is interesting to study the variations of the threshold with the cavity-exciton detuning $\delta$.
Using equations (\ref{hopfield1}), (\ref{hopfield2}) and (\ref{alphapol}) to replace $X_{0}$, $C_{0}$ and
$\alpha_{p}$, the threshold can be written as

\begin{equation}
I_{thr}^{in}= \frac{8\left( \delta ^{2}\gamma _{b}+2\left( \gamma _{a}+\gamma _{b}\right) g^{2}+\delta
\gamma _{b}\sqrt{\delta ^{2}+4g^{2}} \right) ^{3} }{\left({3\sqrt{3}g^{2}\left( \delta +\sqrt{\delta
^{2}+4g^{2}}\right) ^{4}\alpha _{exc} \gamma_{1}} \right)}
\end{equation}

The variations of the threshold versus the cavity-exciton detuning are shown in Fig.~\ref{bistab}. The
threshold intensity is found to reach a minimum value for the detuning $\delta _{0}$ given by

\begin{equation}
\delta _{0}=\frac{2\gamma _{a}-\gamma _{b}}{\sqrt{2\gamma _{a}\gamma _{b}}}g
\end{equation}

With the parameters of Fig.~\ref{bistab} we find $\delta_{0}$=1.72 meV. As explained in the previous
section, the optical response of the system is governed by the composite nature of the polariton, whith
exciton and photon fractions depending on the cavity-exciton detuning. The value $\delta _{0}$ of the
detuning is the result of a trade-off between coupling to the external radiation (which is stronger for
negative detuning, when the polariton tends to a photon) and nonlinearity (which is stronger for positive
detuning, when the polariton tends to an exciton).

\subsection{Reflectivity and transmission spectra}

In this section we compute the reflectivity, absorption and transmission spectra, which show hysteresis
cycles above the bistability threshold.

The reflectivity and transmission coefficients are obtained in the following way. First we compute the
stationary mean value $p_{0}$ of the intracavity polariton field using equation (\ref{intracav2}). Then
we compute the mean value of the intracavity photon field using equations (\ref{def1})-(\ref{def2}) and
the fact that the upper polariton field $q$ is set to zero, which yields the simple relationship
$a=-C_{0}p$. Finally the reflected and transmitted fields are calculated using the input-output
relationships $A_{i}^{out} = \sqrt{2 \gamma_{i}} a - A_{i}^{in}$ for i=1,2. The coefficients $R$, $T$ and
$A$ are given respectively by $I_{1}^{out}/I^{in}$, $I_{2}^{out}/I^{in}$ and 1-$R$-$T$.

\begin{figure}[htbp]
\centerline{\includegraphics[clip=,width=8cm]{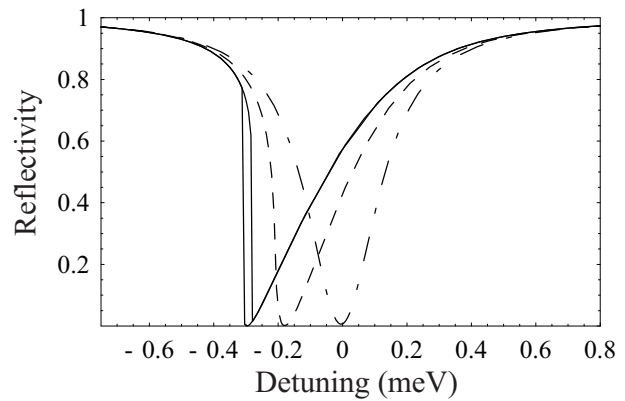}} \caption{Reflectivity spectra as a function of
the cavity-exciton detuning, for $I^{in}$ near zero (dash-dotted line), $I^{in}$=0.5 mW (dashed line) and
$I^{in}$=1 mW (solid line). The laser energy is $E_{L}=E_{exc}-\Omega_{R}/2$, equal to the lower
polariton energy at zero cavity-exciton detuning in the absence of nonlinear effects. The other
parameters are the same as in Fig.~\ref{bistability}. The reflectivity resonance is indeed at $\delta =0$
in the low intensity case but it is shifted at higher intensity. For the highest intensity, a hysteresis
cycle appears when scanning the spot position in the two directions.} \label{reflec1}
\end{figure}

Bistability can be evidenced by scanning the input intensity for a fixed detuning between the exciton and
the cavity. Alternatively, it is possible to scan the cavity length for a fixed value of the input
intensity, as for atoms in cavity \cite{bistabatome}. In a semiconductor microcavity, this can be done by
scanning the excitation spot on the sample surface, since the cavity is wedged (i.e. there is a slight
angle between the Bragg mirrors).

Fig.~\ref{reflec1} shows the variations of R with the cavity-exciton detuning for three values of
${I^{in}}$ (close to zero, below and above the bistability threshold). For the highest intensity, the
reflectivity spectrum shows the characteristic hysteresis cycle. The output power switches abruptly when
the position of the excitation spot is scanned ; in the bistable region the output power depends on the
direction of the scan. The hysteresis cycle can also be seen on the transmission and absorption spectra.

Thus our theoretical model shows that the exciton-exciton interaction in semiconductor microcavities
leads to an optically bistable regime in the frequency region of the polariton resonance. An alternative
mechanism for achieving optical bistability was proposed in Ref.~\cite{tredicucci}, using the bleaching
of the Rabi splitting ; in contrast with this case, we obtain the present effect when the exciton-exciton
interaction term is much smaller than the Rabi splitting term. We also stress that this mechanism is
different from the optical bistability which has been demonstrated in semiconductor microcavities at room
temperature \cite{oudar}, since it involves an exciton-photon mixed mode instead of a cavity mode.

\begin{figure}[htbp]
\centerline{\includegraphics[clip=,width=8cm]{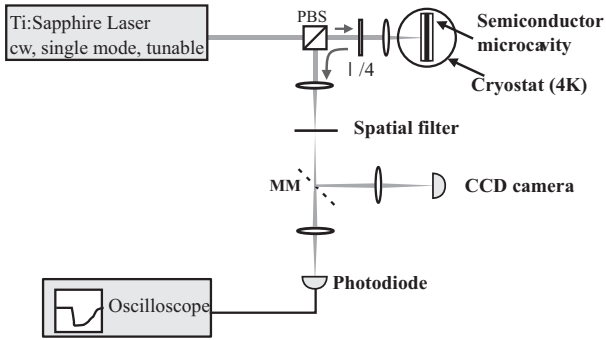}} \caption{Experimental setup. The
microcavity sample is excited using a Ti:Sa laser. The quarter wave plate in front of the sample ensures
excitation with a circular polarization. The polarizing beam splitter (PBS) and the quarter wave plate
form an optical circulator that separate the reflected light from the excitation beam. A spatial filter
is placed in the near field of the reflected beam. Using the movable mirror MM, the beam can be either
observed on a CDD camera, again in the near field (which allows to study the spatial effects and to
choose the position of the spatial filter), or sent towards a photodiode.} \label{montage}
\end{figure}

\section{II. Experimental results}

The microcavity sample consists of one In$_{0.05}$Ga$_{0.95}$As quantum well embedded in a $\lambda$ GaAs
spacer, sandwiched between 20 (26.5) pairs of Ga$_{0.9}$Al$_{0.1}$As/AlAs distributed Bragg reflectors on
top (bottom). The linewidths (FWHM) of the bare exciton and cavity modes are respectively 0.15 meV and
0.24 meV and the Rabi splitting is $\Omega_{R}$=2.8 meV. The sample is held in a cold-finger cryostat at
a temperature of 4K. The cavity has a slight wedge which allows to tune the cavity length by scanning the
position of excitation on the sample. The light source is a single-mode tunable cw Ti:sapphire laser with
a linewidth of the order of 1 MHz. The laser beam is power-stabilized by means of an electro-optic
modulator and spatially filtered by a 2 m-long single-mode fiber. The spot diameter is 50 $\mu$m. In all
experiments the lower polariton branch was excited near resonance at normal incidence with a $\sigma^{+}$
polarized beam.

The image of the excitation spot on the sample surface is made on a CCD camera. Indeed spatial effects
are important in our system. On the one hand, due to the slight angle between the cavity mirrors the
polariton energy depends linearly on the position ; on the other hand the nonlinearity gives rise to
spatial patterns due to the Gaussian intensity distribution in the laser spot. Transverse effects are
thus critical for the understanding of the optical response of the sample and we will present their
experimental study first.

\begin{figure}[htbp]
\centerline{\includegraphics[clip=,width=8.3cm]{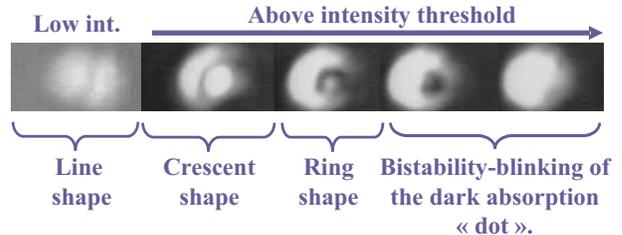}} \caption{Near-field images of the
reflected beam. The laser wavelength is 831.69 nm, resonant with the lower polariton at $\delta$ = 0.3
meV. The first image is taken at very low excitation intensity (0.2 mW). All the other images are taken
at 2 mW, for different positions of the excitation spot on the sample. The last two images are obtained
for the same position ; we observed a blinking between these two states, due to mechanical vibrations.}
\label{imageexp}
\end{figure}

\subsection{Transverse effects}

\subsubsection{Near field images}

We first studied the near field of the reflected beam. At sufficiently low excitation intensity, the spot
shows a dark vertical line corresponding to absorption occurring on the polariton resonance. The
variation of absorption with the position is due to the slight angle between the cavity mirrors, and the
dark line is a line of equal thickness of the microcavity. At higher laser intensities, even well below
the bistability threshold, one observes a strong distortion of the resonance line as shown in
Fig.~\ref{imageexp}. When scanning the spot position on the sample (which amounts to scanning the
polariton energy) one can see a change from a crescent shape to a ring shape, and then to a dot shape.

\begin{figure}[htbp]
\includegraphics[clip=,width=8.3cm]{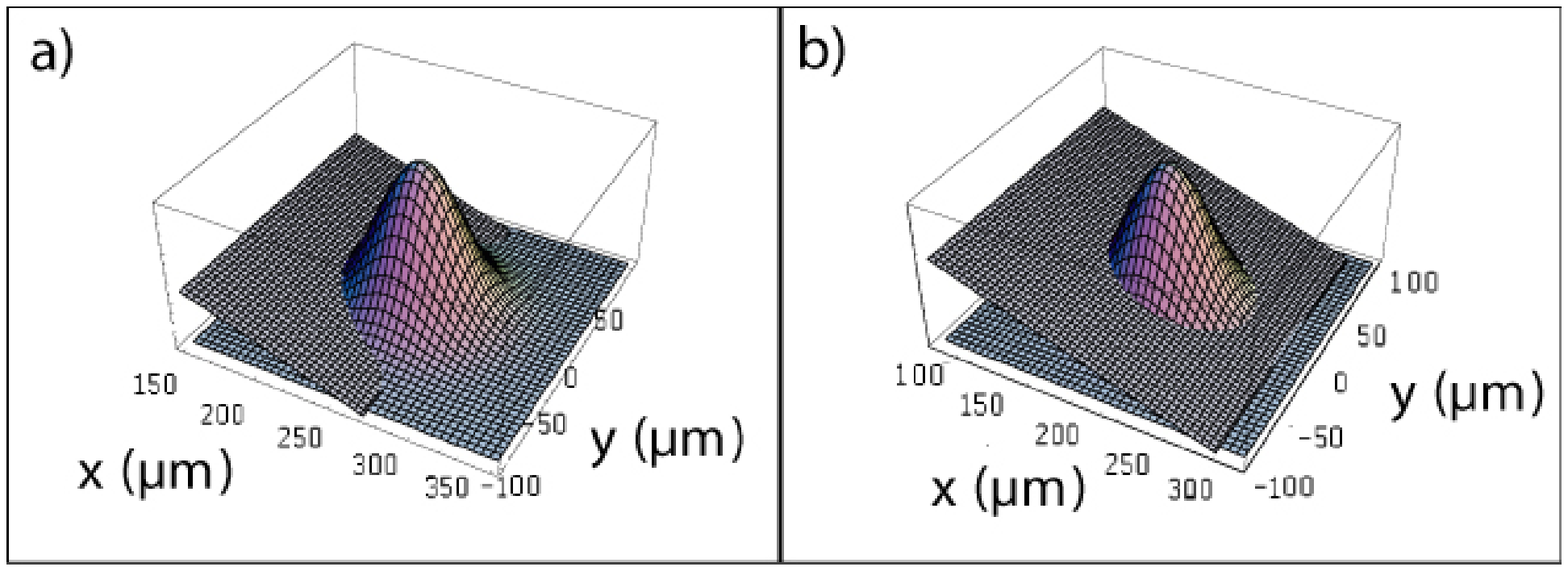} \\
\includegraphics[clip=,width=8.3cm]{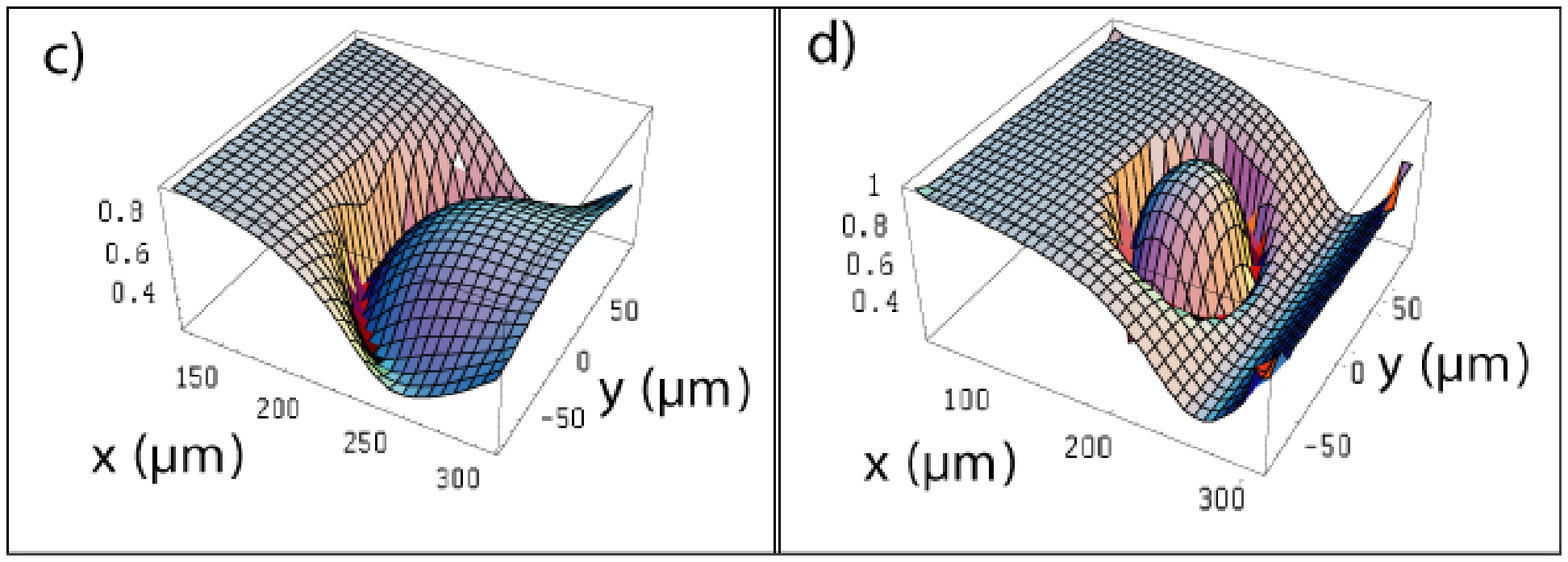} \\
\includegraphics[clip=,width=8.3cm]{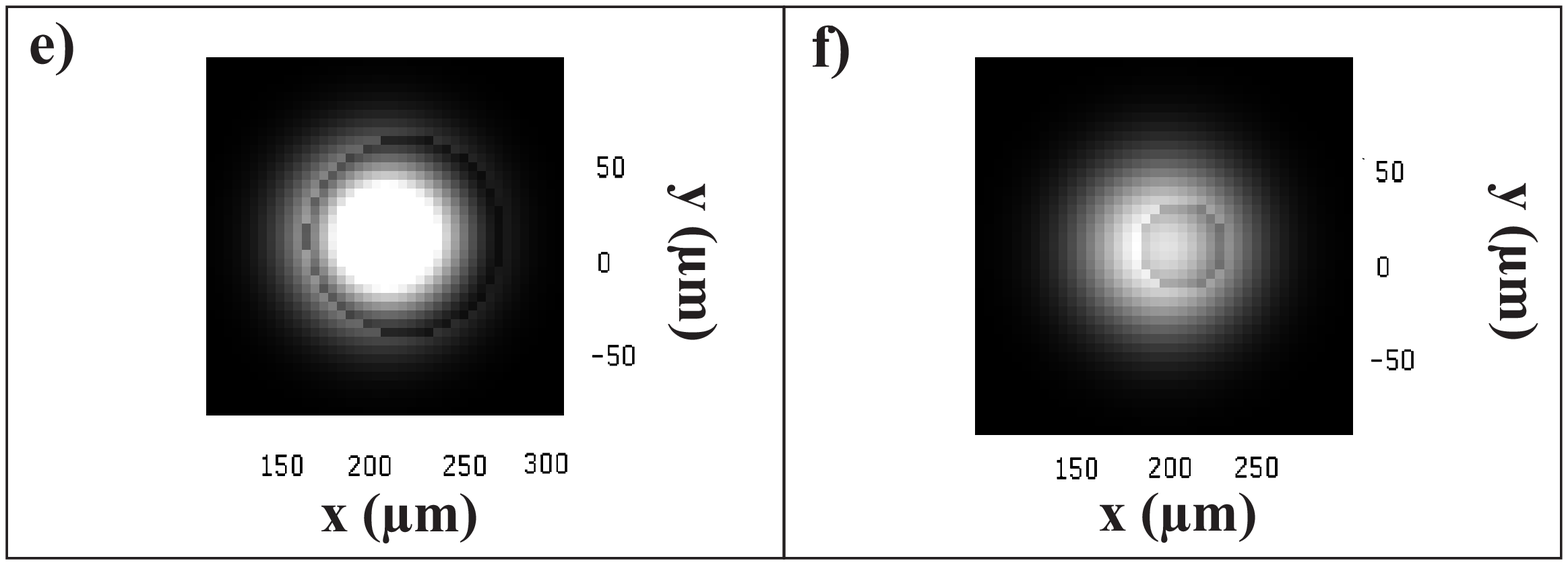}
\caption {All curves are drawn as a function of the position (x, y) in the transverse plane. a) and b):
nonlinear energy shift proportional to the gaussian intensity distribution of the excitation spot and
linear shift due to the cavity wedge, for two positions of the spot on the sample: X=245 $\mu$m et X=180
$\mu$m (X=0 corresponds to zero exciton-cavity detuning). c) and d): reflectivity for the parameters of
a) and b) respectively. The reflectivity resonance is obtained when the nonlinear shift compensates
exactly for the linear shift, i.e. at the intersections between the two curves of Fig.~a) and b). The
low-intensity resonance (a straight line) can be seen on the edge of plot d). The e) and f): near field
images of the spot for the parameters of a) and b) respectively, obtained by convolution of the
reflectivity with the intensity distribution. They are to be compared with the first two pictures in
Fig.~\ref{imageexp}. Note that the low-intensity resonance region of plot d) is not visible in plot f)
since the local intensity is very low. The unshifted resonance is at X=270 $\mu$m.} \label{imagetheo}
\end{figure}

\subsubsection{Theoretical study}

In order to understand these results we have computed the reflectivity of the microcavity in the
transverse plane, taking both the gaussian intensity distribution of the excitation spot and the wedged
shape of the cavity into account. The excitation spot is discretized into small spots with different
local excitation densities and local cavity thicknesses. The pixels are assumed to be uncorrelated to
each other. For each spot we compute the reflectivity from (\ref{evolpolariton}). This is the simplest
possible treatment, which includes neither the transverse mode structure of the microcavity \cite{bjork}
(since it is based on a plane-wave approximation), nor the effect of diffraction. However it gives a good
qualitative understanding of the shape of the absorption region.

At low intensity, the resonance region is found to be a straight line, as observed in the experiments.
The results at higher intensity with the experimental parameters of Fig.~\ref{imageexp} can be seen in
Fig.~\ref{imagetheo} for two different positions of the excitation spot. The main resonant region can be
seen near the center of the spot ; depending on the position of the spot, its shape is that of a
crescent, a ring or a dot, in good agreement with the experimental results. The shape of the resonance
region can be understood as resulting from exact compensation between the nonlinear energy shift due to
the intensity variations, and the linear energy shift due to the cavity thickness variations. The results
of Fig.~\ref{imagetheo} are in good agreement with the first two pictures of Fig.~\ref{imageexp} that
correspond to the same parameters. The blinking of the dark absorption dot as a whole in the bistable
region cannot be reproduced by our model, because it is linked to the spatial coherence over the size of
the dot (i.e. the size of the polariton mode), while in our model there are no spatial correlations
between the pixels used in the calculation.

\begin{figure}[htbp]
\centerline{\includegraphics[clip=,width=8cm]{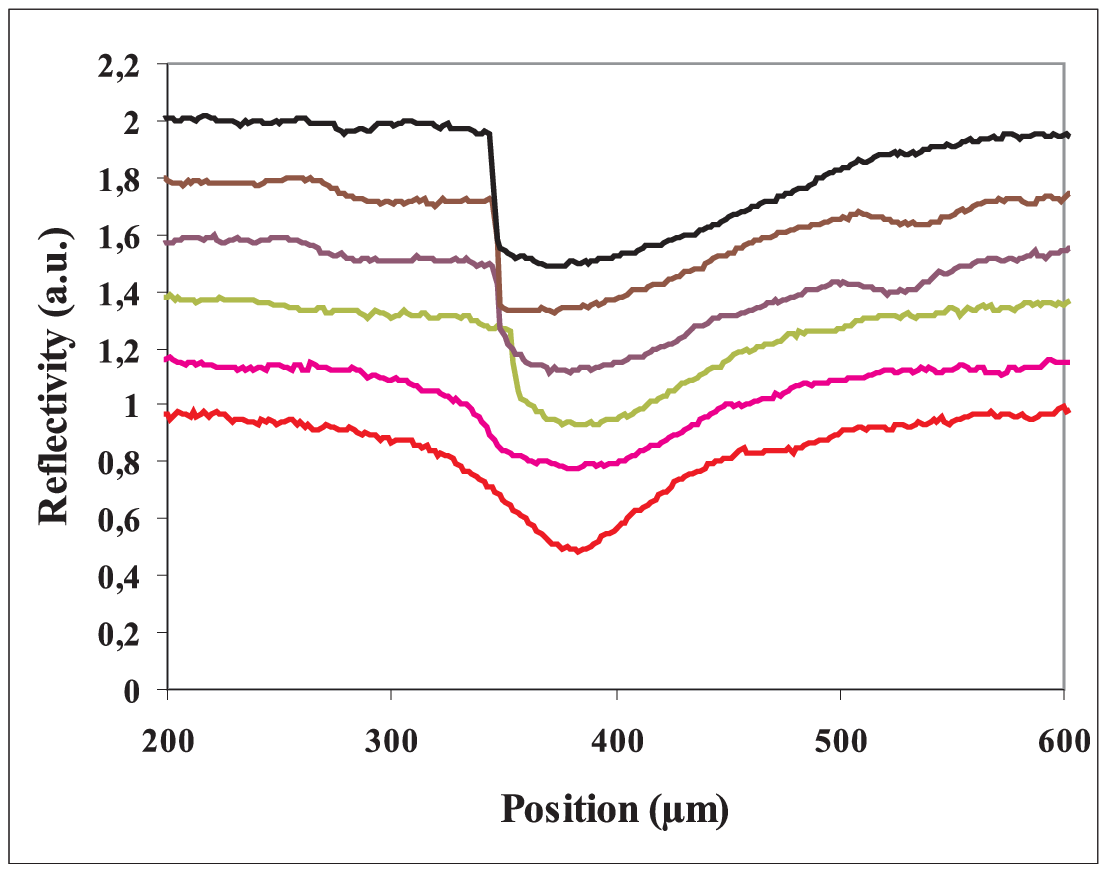}}
\centerline{\includegraphics[clip=,width=8cm]{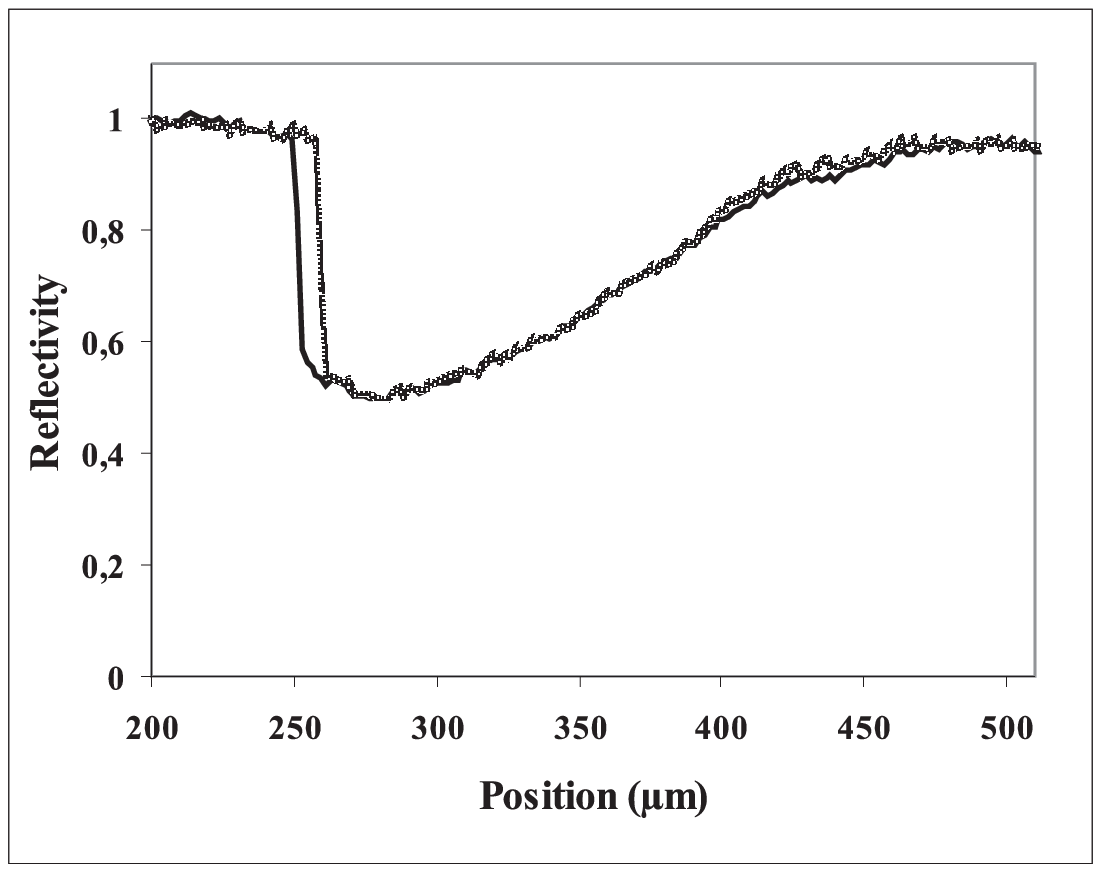}} \caption{Upper figure: reflected
intensities (in arbitrary units) as a function of the spot position on the sample (the origin of the axis
is arbitrary), for several values of the input power $I^{in}$: 1, 2, 3, 4, 5 and 6 mW. The laser
wavelength is 831.32 nm, resonant with the lower polariton at $\delta$ = 1.5 meV. Bistability appears at
$I^{in}$=2.8 mW. Lower figure: hysteresis cycle for the curve $I^{in}$ = 6 mW. The two curves correspond
to the two directions for the scan of the spot position on the sample. } \label{reflecexp}
\end{figure}

\subsection{Reflectivity spectra}

In view of these transverse effects, the interpretation of the reflectivity spectrum will be much simpler
if we select a small zone on the sample in order to avoid averaging the optical response on the spot
surface. One solution would be to have an excitation spot with a uniform distribution in intensity,
sufficiently small so that the cavity wedge would be negligible. Spatial selection can be also easily
achieved by spatial filtering of the reflected beam. We used a spatial filter for the reflected light in
order to select only a small fraction of the excitation spot. The filter has the size of the dark
absorption dot of Fig.~\ref{imageexp}, i.e. about 10 $\mu$m in diameter.

Two photodiodes allow to measure the reflected and transmitted intensities. Each spectrum is obtained at
fixed excitation energy and intensity, by scanning the spot position over the sample surface. This is
equivalent to a scan of the cavity length. As a result, the precision of the measurements is limited by
the surface defects of the sample. Fig.~\ref{reflecexp} shows a series of spectra for several values of
the excitation intensity.

\begin{figure}[htbp]
\centerline{\includegraphics[clip=,width=8cm]{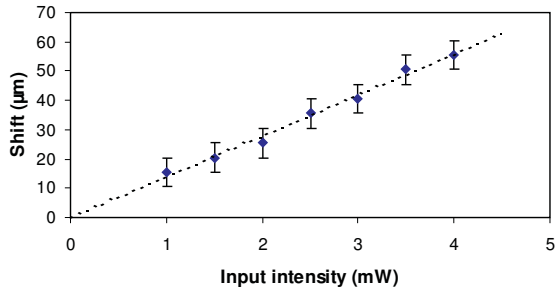}} \caption{Shift of the reflectivity minimum on
the sample, as a function of the input power. The laser wavelength is 831.32 nm, resonant with the lower
polariton at $\delta$ = 1.5 meV.} \label{shift}
\end{figure}

We observed a shift of the resonance position, which is proportional to the excitation intensity (see
Fig.~\ref{shift}). The resonance position shifts towards negative detunings, corresponding to a blueshift
of the resonance energy. This is in agreement with the model, where the blueshift is given by the
effective Hamiltonian (\ref{nonlineaire}). The shift has been removed in Fig.~\ref{reflecexp}, so that
all the curves appear to be peaked around the same position.

Above a threshold intensity, one can observe as expected an hysteresis cycle by scanning the sample
position in the two directions (Fig.~\ref{reflecexp}). The bistability threshold can be determined with a
good precision by observing the spontaneous "blinking" between the two stable values, due to the
intensity fluctuations or mechanical vibrations of optic elements on the set-up. In the case of
Fig.~\ref{reflecexp} $I^{in}_{thr}$ = 2.8 mW.

\begin{figure}[htbp]
\centerline{\includegraphics[clip=,width=8cm]{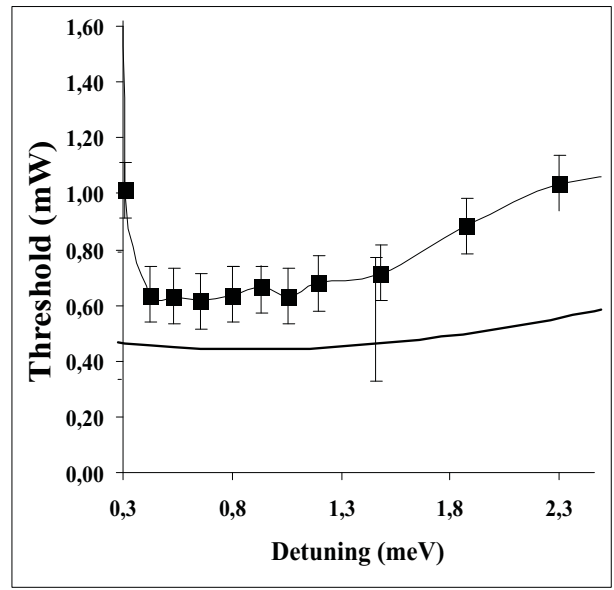}} \caption{Bistability threshold versus
exciton-cavity detuning. Squares: experimental data (the line is a guide for the eyes). Solid line:
theoretical curve calculated from the measured values of $d_{spot}$ = 38 $\mu$m and $\gamma_{p}$ = 0.125
meV. We added an error bar on the point at $\delta$ = 1.5 meV of the theoretical curve showing the
uncertainty on the theoretical value coming from the uncertainty in the measurement of the spot size and
the polariton linewidth.} \label{seuilexp}
\end{figure}

\subsubsection{Bistability threshold}

The variations of the bistability threshold with the cavity-exciton detuning are shown in
Fig.~\ref{seuilexp}. We didn't use any spatial filter here.

In the context of our previous assumption of small independent pixels, a precise calculation of the
bistability threshold is not possible ; the model predicts that each pixel of the excitation spot has its
own bistability threshold. However, we can make a crude estimate by assuming that bistability occurs as
soon as the local density of excitation in the center of the spot exceeds the bistability threshold
calculated in (\ref{threshold}). As a result, we have to divide the threshold (\ref{threshold}) by the
ratio of the local density at the center of the spot by the mean density of excitation, which is found to
be $\sqrt{2}$.

The value of the nonlinear coefficient $\alpha_{p}$ is calculated using expression (\ref{alphapol}) with
the measured spot diameter of $d_{spot}$ = 38 $\pm$4 $\mu$m. The polariton linewidth $\gamma_{p}$ is
estimated from a reflectivity measurement at $\delta$ = 1.5 meV, which must be taken at threshold because
the linewidth increases with the excitation density due to collision broadening. For the sake of
simplicity, we have neglected the variations of $\gamma_{p}$ in the considered range of cavity-exciton
detunings (inferior to ten percent). The solid line in Fig.~\ref{seuilexp} is the result of the model. It
is in fair agreement with the experimental curve. As an illustration, error bars have been added for the
point on the theoretical curve at $\delta$ = 1.5 meV corresponding to the uncertainties in measuring the
parameters of the fit $\alpha_{exc}$ and $\gamma_{p}$.

The general shape of the curve corresponds to the expected behavior, except when the detuning goes down
to zero where the threshold intensity goes up very quickly (for example it is superior to 25 mW for
$\delta$=0.3 meV). A more elaborate model including the transverse mode structure of the microcavity and
diffraction is needed in order to understand this effect. The minimal threshold is obtained at slightly
positive detuning, near the theoretical value.

\section{Conclusion}

We have reported the first observation of polaritonic bistability in semiconductor microcavities. It is
obtained in the context of parametric polariton interaction in the degenerate case, where only the pumped
mode at $k=0$ is involved. A simple model treating the non-linearity as a Kerr type one gives a
relatively good agreement with the experimental results. The originality of the bistable behavior
reported here, is that it occurs in the strong coupling regime: the effective refraction index depends on
the polariton number instead of the photons number. Moreover, it is well-known that Kerr media can be
used for the production of squeezed states of light. In the same way, this non-linear mechanism can be
used to produce squeezed states of the mixed light-matter polariton field \cite{karr03} \footnote{This
model will be extended in a forthcoming paper to include quantum fluctuations in order to understand how
squeezing can be achieved.}.

\section{Acknowledgements}
We thank Romuald Houdr\'{e} for providing us with the microcavity sample.

\end{document}